\documentclass[11pt]{article}

\usepackage{amssymb}
\usepackage{amsmath}
\usepackage{mathpazo}
\usepackage{color}
\usepackage{graphicx}
\usepackage[mathscr]{eucal}
\usepackage[colorlinks,linkcolor=blue]{hyperref}

\def\calB{{\cal B}}

\def\calG{{\cal G}}

\def\sfA{{\mathsf{A}}}

\def\sfC{{\mathsf{C}}}
\def\sfd{{\mathsf{d}}}
\def\sfD{{\mathsf{D}}}

\def\sfE{{\mathsf{E}}}

\def\sfF{{\mathsf{F}}}
\def\sfg{{\mathsf{g}}}

\def\sfQ{{\mathsf{Q}}}

\def\sfR{{\mathsf{R}}}
\def\sfs{{\mathsf{s}}}

\def\sft{{\mathsf{t}}}

\def\sfv{{\mathsf{v}}}
\def\sfw{{\mathsf{w}}}

\def\sfZ{{\mathsf{Z}}}

\def\bfA{{\mathbf{A}}}

\def\bfC{{\mathbf{C}}}
\def\bfd{{\mathbf{d}}}
\def\bfD{{\mathbf{D}}}
\def\bfe{{\mathbf{e}}}

\def\bfI{{\mathbf{I}}}
\def\bfg{{\mathbf{g}}}
\def\bfG{{\mathbf{G}}}

\def\bfm{{\mathbf{m}}}
\def\bfM{{\mathbf{M}}}

\def\bfQ{{\mathbf{Q}}}
\def\bfR{{\mathbf{R}}}

\def\bfs{{\mathbf{s}}}
\def\bfS{{\mathbf{S}}}

\def\bfv{{\mathbf{v}}}
\def\bfw{{\mathbf{w}}}
\def\bfx{{\mathbf{x}}}
\def\bfy{{\mathbf{y}}}

\def\SpaceV{{\mathbb{V}}}
\def\SpaceW{{\mathbb{W}}}
\def\SpaceX{{\mathbb{X}}}
\def\SpaceY{{\mathbb{Y}}}

\newcommand{\sfDelta}{\mbox{\textsf{$\Delta$}}}

\newcommand{\sfSigma}{\mbox{\textsf{$\Sigma$}}}

\newcommand{\sfomega}{\mbox{\textsf{$\omega$}}}

\newcommand{\bfomega}{\mbox{\boldmath$\omega$}}

\newcommand{\bfOmega}{\mbox{\boldmath$\Omega$}}

\def\PointP{{\mathscr{P}}}
\def\PointM{{\mathscr{M}}}
\def\bfE{{\bf E}}

\title{Stress and strain in\\
       symmetric and asymmetric\\
       elasticity}
\author{Albert Tarantola\thanks{Universit\'e de Paris VI \& Institut de Physique du Globe de Paris, 4 place Jussieu, 75005~Paris, \,France \,(albert.tarantola@ipgp.fr).}}

\begin{document}

\maketitle

%
% abstract
%
\begin{abstract}
Usual introductions of the concept of motion are not well adapted to a subsequent, strictly tensorial, theory of elasticity. The consideration of arbitrary coordinate systems for the representation of both, the points in the laboratory, and the material points (comoving coordinates), allows to develop a simple, old fashioned theory, where only measurable quantities ---like the Cauchy stress--- need be introduced. The theory accounts for the possibility of asymmetric stress (Cosserat elastic media), but, contrary to usual developments of the theory, the basic variable is not a micro-rotation, but the more fundamental micro-rotation \emph{velocity}. The deformation tensor here introduced is the proper tensorial equivalent of the poorly defined deformation ``tensors'' of the usual theory. It is related to the deformation velocity tensor via the matricant. The strain is the logarithm of the deformation tensor. As the theory accounts for general Cosserat media, the strain is not necessarily symmetric. Hooke's law can be properly introduced in the material coordinates (as the stiffness is a function of the material point). A particularity of the theory is that the components of the stiffness tensor in the material (comoving) coordinates are not time-dependent. The configuration space is identified to the part of the Lie group \,GL$^+$(3) that is geodesically connected to the origin of the group.
\end{abstract}

\newpage

%
% table of contents
%

\tableofcontents

%\newpage

%
% section
%
\section{Introduction}

Elasticity has often been the model theory for building other theories. For instance, Maxwell used the elastic analogy to develop his equations, and \'Elie Cartan was inspired by elasticity\footnote{Cosserat theory, with asymmetric stress.} to propose the Einstein-Cartan gravitation theory\footnote{While in Einstein's theory space-time may have curvature but no torsion, in the Eistein-Cartan theory, both curvature and torsion may exist, torsion allowing to take into account the existence of spin in realistic models of matter.}. 

Today, elasticity is mainly being developed by applied mathematicians, whose principal goal is mathematical generality and consistency, at the cost ---in my opinion--- of blurring the difference between the tensors beloved by Cauchy and Einstein\footnote{A tensor is an intrinsically defined quantity at some point of the differentiable manifold representing the space-time. It belongs to one of the linear spaces that can be built by different tensor products of the tangent linear manifold and its dual. The sum and the product by a real number are the two basic operations for tensors. In reality, there are objects ---like a rotation operator--- that share with tensors the property of being intrinsically defined, but whose natural operations are the product and the exponentiation to a real number. These are not ---strictly speaking--- tensors, and are better seen as points on some Lie group tangent to the space-time manifold (Tarantola, 2006). Their logarithm is almost a tensor, as the sum and product for a real number is defined, but this sum is not commutative. So the reader may now understand how restrictive this author is in the use of the term {\sl tensor}\label{test}.}, and their generalization as abstract mappings between linear spaces (subjected to pull backs and pull forwards). Also, some confusion exists on the physical interpretation of the the different stress ``tensors'' introduced in the usual theory, that are better seen as just auxiliary computational tools. Therefore, in this note I take the notion of tensor much more basically than usual texts in deformation theory. And I resolutely take the old-fashioned index-based notation: it is always easy, once the basic mathematics and physics are understood, to move towards more abstract notations.

Because I want to highlight some simple messages, I completely ignore the dynamic part of the problem (conservation of mass, of linear momentum, and of angular momentum are not considered): only the definition of strain, of stress, and of internal elastic energy are examined. And, of course, the relation between stress and strain.

We are now celebrating the one-hundredth anniversary of the ground-breaking work of the Cosserat brothers (Cosserat and Cosserat, 1909). Most of the scientists familiar with the Cosserat theory understand that the usual assumption of symmetric stress breaks the beauty of the mathematics, in exactly the same way as anyone who understands the Einstein-Cartan theory of gravitation knows that assuming vanishing space-time torsion (thus, a symmetric connection) takes out most of the beauty of the Bianchi conservation equations.

So, in this theory, the stress tensor can be asymmetric. And, following the Cosserats, we interpret the antisymmetric part of the stress as the effect of micro-rotations (of, say, the material ``molecules'').
Yet, I choose not to follow common practice of explicitly considering the micro-rotations as a primitive ingredient of the theory. For a rotation is always relative to some initial configuration, whereas a {\emph{rotation velocity} is not. And, for the same reason, (symmetric) strain is not a primitive ingredient; the \emph{deformation velocity} is. This matters, as the deformation velocity is not simply the strain rate.

%
% section
%
\section{Movement}

%
% section
%
\subsection{Frame of reference}

Let us start by assuming the existence of Galilean frames of reference, and by choosing a particular one, say \,$ \calG $\,, with respect to which all tensor fields (velocities, stresses, etc.) are defined. The time coordinate \,$ t $\, is Newtonian time. We do not need to assume that the spatial part of the Galilean frame~\,$ \calG $\, is Euclidean\footnote{What is convenient if the theory is to be applied to some non-flat submanifold of the three-dimensional physical space.}, or that it is necessarily thee-dimensional (although I shall use a language adapted to the three-di\-men\-sio\-nal case). A~space point of \,$ \calG $\, may be denoted using a letter like~\,$ \PointP $\,.

A \emph{tensor field} may be represented using a notation like
\begin{equation}
\sfs \ = \ \sigma(\PointP,t) \quad .
\label{frist-abstrat-12345}
\end{equation} 
Here, \,$ \sfs $\, denotes the tensor at space-time point \,$ \{\PointP,t\} $\, while \,$ \sigma $\, denotes the \emph{function} of the space-time coordinates, exactly as when ---in elementary mathematics--- one writes \,$ y \, = \, f(x) $\,.

%
% section
%
\subsection{Motion}

A deforming body \,$ \calB $\, is assumed to occupy the whole\footnote{The modifications to be made when the body \,$ \calB $\, occupies only part of the space are quite trivial conceptually, although technically complex.} of the space. Its points, called \emph{material points}, are assumed to be individually identifiable and their \emph{trajectory} in the Galilean frame \,$ \calG $\, knowable (at least, in principle). The trajectories of the material points is the function
\begin{equation}
\PointP \ = \ f(\PointM,t) \quad ,
\label{trajectories_1234567}
\end{equation}
specifying, at every instant \,$ t $\,, the laboratory position \,$ \PointP $\, of any material point \,$ \PointM $\,. The continuity hypothesis is that the inverse function exist:
\begin{equation}
\PointM \ = \ F(\PointP,t) \quad .
\end{equation}
With this, equation~\eqref{frist-abstrat-12345} can be completed by introducing a new function:
\begin{equation}
\begin{split}
\sfs \ 
& = \ \sigma(\PointP,t) \qquad \qquad ( \, \PointP \ = \ f(\PointM,t) \, ) \\
& = \ \Sigma(\PointM,t) \hskip 13.8 mm ( \, \PointM \ = \ F(\PointP,t) \, ) \quad . \\
\end{split}
\label{frist-abstrat-12345rr}
\end{equation} 

%
% section
%
\subsection{Laboratory coordinates}

The theory could be developed using intrinsic notations only, i.e., without writing tensor equations in terms of the components of the tensors in the natural basis associated to some coordinates. But it is well-known that, as far as the coordinates are arbitrary, component-based tensor equations \emph{are} intrinsic. Many mathematicians prefer more abstract, component-free, notations, and there is no problem with that, excepted that pedagogy may command using expressions that are as explicit as possible. Sometimes, one fails to understand that coordinates are a tool for discovery: there are many examples where it is the discovery of a system of coordinates adapted to a problem that has allowed its resolution\footnote{The discovery of the Schwarzschild coordinates in space-time was fundamental for the discovery of spherically symmetrical solutions to Einstein's equations. More recently, it was the discovery of a space-time coordinate system that allowed to properly formulate ---and solve--- the problem of a fully relativistic positioning system (Coll, 2002; Tarantola et al., 2009). Not to speak about Lagrange's discovery of the material coordinates\dots}.

So, let us assume that some (fixed, arbitrary) coordinate system, \,$ x \, = \, \{x^i\} \, = \, \{x^1,x^2,x^3\} $\, is chosen in the spatial part of the Galilean frame of reference\footnote{The space is not assumed to be Euclidean, so, a fortiori, these coordinates are not assumed to be Cartesian.}. This coordinate system, together with the Newtonian time \,$ t $\, constitute what we shall call a system of (space-time) \emph{laboratory coordinates}. At every space point \,$ x $\, the natural vector basis \,$ \bfe_{i}(x) $\, is considered, together with the associated natural tensor basis \,$ \bfe_i(x) \otimes \bfe_j(x) \dots \bfe^k(x) \otimes \bfe^\ell(x) \dots $\,. Then, for the tensor field in equation~\eqref{frist-abstrat-12345} we can now write:
\begin{equation}
\begin{split}
\sfs \ 
& = \ \bfs(x,t) \ = \ s_{ij\dots}{}^{k\ell\dots}(x,t) \ \bfe^i(x) \otimes \bfe^j(x) \dots \bfe_k(x) \otimes \bfe_\ell(x) \dots\quad . \\
\end{split}
\label{frist-abstrat-12346}
\end{equation}
Note that, by definition of the laboratory coordinates, the vector basis \,$ \bfe_i(x) $\, is time-independent (contrary to the comoving vector basis about to be introduced).

%
% section
%
\subsection{Material (comoving) coordinates}

Let us now assume that an arbitrary system of \emph{material coordinates} \,$ X \, = \, \{ X^I \} \, = \, \{ X^1,X^2,X^3\} $\, has been chosen, as suggested in figure~\ref{Deformation}. By definition, the material coordinates of any material point \,$ \PointM $\, have constant values. The trajectories in equation~\eqref{trajectories_1234567} have now the more concrete, coordinate-based, representation
\,$ x^i \, = \, \phi^i(X^1,X^2,X^3,t) \ , \ (i = 1,2,3) $\,, or, for short,
\begin{equation}
x \ = \ \phi(X,t)  \quad .
\label{trajectories_1234569}
\end{equation}
The \emph{continuity hypothesis} is that these functions are continuous and invertible, so that the inverse functions exist:
\,$ X^I \, = \, \Phi^I(x^1,x^2,x^3,t) \ , \  (I = 1,2,3) $\,. For short, we simply write
\begin{equation}
X \ = \ \Phi(x,t)  \quad .
\label{trajectories_1234570}
\end{equation}

Seen from the laboratory (i.e., from the Galilean frame of reference), the material coordinates deform. Therefore, the natural vector basis associated to the material coordinates (these vectors ---as all tensors of the theory--- are defined with respect to \,$ \calG $\,) is time-dependent, so the notation \,$ \bfe_I(X,t) $\, has to be used. The availability of the material system of coordinates, and of the associated tensor basis, allows to further complete equation~\eqref{frist-abstrat-12346}, introducing also the material components of the tensor field:
\begin{equation}
\begin{split}
\sfs \ 
& = \ \bfs(x,t) \\
& = \ s_{ij\dots}{}^{k\ell\dots}(x,t) \ \bfe^i(x) \otimes \bfe^j(x) \dots \bfe_k(x) \otimes \bfe_\ell(x) \dots\\
& = \ \bfS(X,t) \\
& = \ s_{IJ\dots}{}^{KL\dots}(X,t) \ \bfe^I(X,t) \otimes \bfe^J(X,t) \dots \bfe_K(X,t) \otimes \bfe_L(X,t) \dots \quad , \\
\end{split}
\label{frist-abstrat-12347}
\end{equation}
with the understanding that for these identities to make sense one has to use the replacements in equations~\eqref{trajectories_1234569}--\eqref{trajectories_1234570}.

%
% figure
%
\begin{figure}[htbp]
   \centering
   \includegraphics[width=\textwidth]{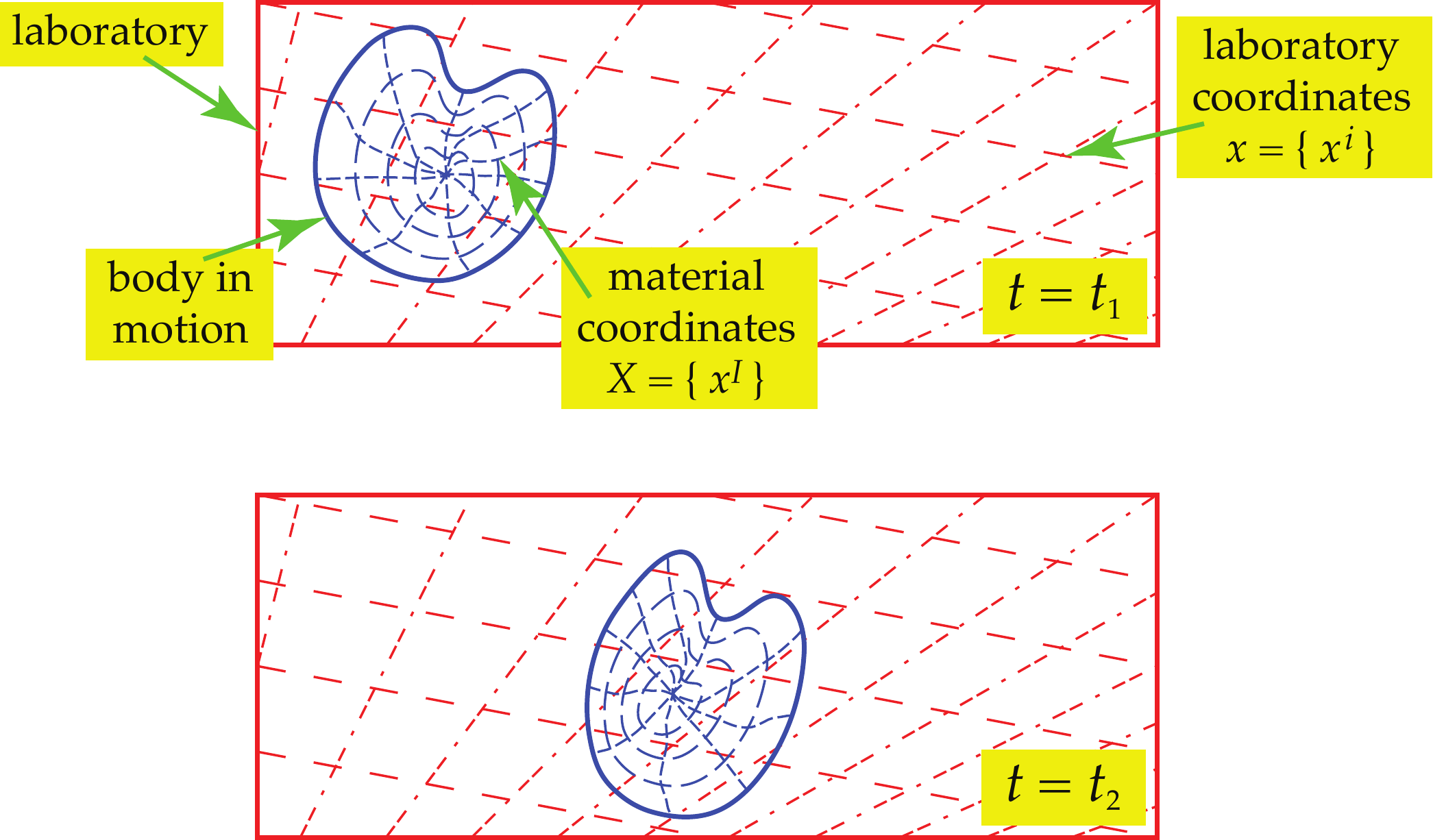} 
   \caption{The ``laboratory'' is assumed to be a Galilean frame of reference. Two snapshots of the laboratory, at two instants \,$ t_1 $\, and \,$ t_2 $\,, with a deforming body occupying part of the laboratory space. In red, a system of laboratory coordinates, \,$ x \, = \, \{ x^i \} $\,, and, in blue, a material (or comoving) system of coordinates, \,$ X \, = \, \{ x^I \} $\,. The physical space is not necessarily Euclidean, and both coordinate systems are arbitrary. The motion of the body is represented by the functions \,$ x \, = \, \phi(X,t) $\,, prescribing, for any time~\,$ t $\,, the laboratory position \,$ x $\, of every material point point \,$ X $\,. This author believes that the usual drawings, with plenty of axes and vectors, involve superfluous notions.}
   \label{Deformation}
\end{figure}

%
% section
%
\subsection{Coordinate change}

Associated to the coordinate changes in equations~~\eqref{trajectories_1234569}--\eqref{trajectories_1234570} are the matrices of coefficients
\begin{equation}
\lambda^i{}_I(X,t) \ = \ \frac{\partial \phi^i}{\partial X^I}(X,t) \qquad ; \qquad 
\lambda^I{}_i(x,t) \ = \ \frac{\partial \Phi^I}{\partial x^i}(x,t) \quad ,
\end{equation}
that are mutually inverse:
\,$ \lambda^i{}_K \, \lambda^K{}_j \, = \, \delta^i{}_j $\,, 
\,$ \lambda^I{}_k \, \lambda^k{}_J \, = \, \delta^I{}_J $\,.
A well-known result from tensor calculus is that the vector and form bases in equation~\eqref{frist-abstrat-12347} are related as\footnote{I.e., more explicitly
\,$ \bfe_I(X,t) = \lambda^i{}_I(X,t) \, \bfe_i( \, \phi(X,t) \,) $\,,
\,$ \bfe^I(X,t) = \lambda^I{}_i( \, \phi(X,t) \, , \, t \, ) $ $ \bfe^i( \, \phi(X,t) \,) $\,,
\,$ \bfe_i(x) = \lambda^I{}_i(x) \, \bfe_I( \, \Phi(x,t) \, , \, t \, ) $\,,
\,$ \bfe^i(x) = \lambda^i{}_I( \, \Phi(x,t) \, , \, t \, ) \, \bfe^I( \, \Phi(x,t) \, , \, t \, ) $\,.
} 
\,$ \bfe_I = \lambda^i{}_I \, \bfe_i $\,,
\,$ \bfe^I = \lambda^I{}_i \, \bfe^i $\,,
\,$ \bfe_i = \lambda^I{}_i \, \bfe_I $\,,
\,$ \bfe^i = \lambda^i{}_I \, \bfe^I $\,,
from where follows that the components \,$ s_{ij\dots}{}^{k\ell\dots}(x,t) $\, are related to the components \,$ s_{IJ\dots}{}^{KL\dots}(X,t) $\, as\footnote{I.e., more explicitly,
\,$ s_{IJ\dots}{}^{KL\dots}(X,t) \ = \ \lambda^i{}_I(X,t) \, \lambda^j{}_J(X,t) \, \dots \, \lambda^K{}_k( \, \Phi(x,t) \, , \, t \, ) \, \lambda^L{}_\ell( \, \Phi(x, $ $ t) \, , \, t \, ) \, \dots \, s_{ij\dots}{}^{k\ell\dots}( \, \Phi(x,t) \, , \, t \, ) $\,.
}
\begin{equation}
s_{IJ\dots}{}^{KL\dots} \ = \ \lambda^i{}_I \, \lambda^j{}_J \, \dots \, \lambda^K{}_k \, \lambda^L{}_\ell \, \dots \, s_{ij\dots}{}^{k\ell\dots} \quad ,
\end{equation}
or, equivalently, as\footnote{I.e., more explicitly,
\,$ s_{ij\dots}{}^{k\ell\dots}(x,t) \ = \ \lambda^I{}_i(x,t) \, \lambda^J{}_j(x,t) \, \dots \, \lambda^k{}_K ( \, \Phi(x,t) \, , \, t \, ) \, \lambda^\ell{}_L( \, \Phi(x,t) \, , $ $ t \, ) \, \dots \, s_{IJ\dots}{}^{KL\dots}( \, \Phi(x,t) \, , \, t \, ) $\,.
}
\begin{equation}
s_{ij\dots}{}^{k\ell\dots} \ = \ \lambda^I{}_i \, \lambda^J{}_j \, \dots \, \lambda^k{}_K \, \lambda^\ell{}_L \, \dots \, s_{IJ\dots}{}^{KL\dots} \quad .
\end{equation}

%
% section
%
\subsection{Metric}

We shall assume that the components of the metric \,$ g_{ij}(x) $\, are known in the laboratory coordinates. Quite often, the space is going to be Euclidean, and, in this case, the \,$ g_{ij}(x) $\, are just the components of the Euclidean metric in the (arbitrary) laboratory coordinates, but let us \emph{not} assume that we are in this special situation.
We can naturally write
\begin{equation}
\sfg \ = \ \bfg(x) \ = \ g_{ij}(x) \ \bfe^i(x) \otimes \bfe^j(x) \quad ,
\end{equation}
denoting by the symbol \,$ \sfg $\, the metric tensor itself.
Note that the components \,$ g_{ij}(x) $\, of the metric tensor in the laboratory coordinates are time-independent. This is not so in the material coordinates, where one has
\begin{equation}
\sfg \ = \ \bfG(X) \ = \ g_{IJ}(X,t) \ \bfe^I(X,t) \otimes \bfe^J(X,t) \quad ,
\label{the g to derive}
\end{equation}
as both, the basis \,$ \bfe_I(X,t) $\, and the components \,$ g_{IJ}(X,t) $\,, are time-depen\-dent. The components \,$ g_{IJ} $\, can be expressed as a function of the components \,$ g_{ij} $\, as
\,$ g_{IJ} \, = \, \lambda^i{}_I \, \lambda^j{}_J \, g_{ij} $\,,
i.e., more explicitly,
\begin{equation}
g_{IJ}(X,t) \ = \ \lambda^i{}_I(X,t) \, \lambda^j{}_J(X,t) \, g_{ij}( \, \phi(X,t) \, ) \quad .
\end{equation}

%
% section
%
\subsection{Velocity}

The considered motion defines a velocity field, namely, the velocity of all the material points with respect to the Galilean frame of reference,
\begin{equation}
v^i\ = \ \partial \phi^i / \partial t \quad , 
\label{velofield-458}
\end{equation}
an equation that may be better understood if all the variables are explicited:
\,$ v^i(x,t) \, = \, (\partial \phi^i/\partial t)(\, \Phi(x,t) \, , \, t \, ) $\,.
To obtain the expression of the velocity field in the material coordinates, one can just use
\,$ v^I \, = \, \lambda^I{}_i \, v^i $\,. Alternatively, one has 
\begin{equation}
v^I \ = \ - \, \partial \Phi^I / \partial t \quad ,
\end{equation}
i.e.,
\,$ v^I(X,t) \ = \, - \, (\partial \Phi^I/\partial t)( \, \phi(X,t) \, , \, t \, ) $\,.
The covariant components of the vector field are
\,$ v_i(x,t) \, = \, g_{ij}(x) \, v^j(x,t) $\, and
\,$ v_I(X,t) \, = \, g_{IJ}(X,t) \, v^J(X,t) $\,.

%
% section
%
\subsection{Deformation velocity}

The notion of ``strain tensor'' is subtle, and only to be introduced later. A~robust notion is that of \emph{deformation velocity} (similar, but different from the ``strain rate'' to be later introduced). The deformation velocity tensor, denoted \,$ {\sf d} $\,, can be introduced using any of the two equivalent definitions
\begin{equation}
d_{IJ} \ = \ \tfrac{1}{2} \, (\nabla_{\!I} v_J + \nabla_{\!J} v_I ) \qquad ; \qquad
d_{ij} \ = \ \tfrac{1}{2} \, (\nabla_{\!i} v_j + \nabla_{\!j} v_i ) \quad . 
\label{defo-velo-448427}
\end{equation}
Because of the definition of material coordinates, one has the property\footnote{This can be obtained by evaluating the partial time derivative of expression~\eqref{the g to derive}. As \,$ \partial_t \bfG \, = \, 0 $\,, the result follows when using the property \,$ \partial_t \bfe^I \, = \, - (\nabla_{\!K} v^I) \, \bfe^K $\,.}
\begin{equation}
\boxed{ \qquad 
d_{IJ}(X,t) \ = \ \tfrac{1}{2} \, \dot{g}_{IJ}(X,t) \quad . \quad }
\label{Jeroen formula 1234}
\end{equation}

The \emph{vorticity}
\begin{equation}
\varpi_{IJ} \ = \ \tfrac{1}{2} \, (\nabla_{\!I} v_J - \nabla_{\!J} v_I ) \qquad ; \qquad
\varpi_{ij} \ = \ \tfrac{1}{2} \, (\nabla_{\!i} v_j - \nabla_{\!j} v_i ) \quad ,
\label{vorticity_9945}
\end{equation}
represents a local ``mesoscopic rotation velocity''. It has not to be mistaken for the fundamental ``microscopic rotation velocity'', about to be introduced.

%
% section
%
\subsection{Rotation velocity}

So far, the considered motion has only considered the ``translational'' movements of the material points, considered as featureless. More realistic continuous models of matter also consider the possibility that the individual material points (i.e., the ``molecules'') can rotate. This is, for example, the case for the fluids where the existence of a spin density is to be considered\footnote{For the theoretical beauty of a relativistic theory of fluids with spin, see Halwachs (1960).}. It is also the case in the theory of elastic media where the stress tensor is not assumed to be symmetric. This theory, developed by Eug\`ene and Fran\c{c}ois Cosserat (Cosserat and Cosserat, 1909), also contains micro-rotations. 

My goal in this note is the study of general elastic media, not of fluids with spin. But I think it is a mistake to develop the theory of asymmetric elasticity starting with the notion of rotation. For the notion of (instantaneous) \emph{rotation velocity} is more primitive. If necessary, the rotations have to be evaluated by properly integrating the rotation velocity.

So, let us consider (Cosserat media) that the material points (or ``mo\-le\-cu\-les''), in addition to their translational motion, may rotate, i.e., every material point shall have associated a rotation velocity. This corresponds to an antisymmetric tensor field:
\begin{equation}
\begin{split}
\sfomega \ 
& = \ \bfOmega(X,t) \ = \ \omega_{IJ}(X,t) \ \bfe^I(X,t) \otimes \bfe^J(X,t) \\
& = \ \bfomega(x,t) \ = \ \omega_{ij}(x,t) \ \bfe^i(x) \otimes \bfe^j(x) \quad , \\
\end{split}
\end{equation}
with \,$ \omega_{IJ} + \omega_{JI} = 0 $\, and \,$ \omega_{ij} + \omega_{ji} = 0 $\,.
Again, this intrinsic (or ``microscopic) rotation is not to be mistaken for the vorticity (equation~\eqref{vorticity_9945}).

%
% section
%
\subsection{Movement velocity}

The deformation velocity tensor \,$ \sfd $\, is, by definition, symmetric. The sum of the symmetric deformation velocity and of the antisymmetric rotation velocity,
\begin{equation}
\Delta_{IJ} \ = \ d_{IJ} + \omega_{IJ} \qquad ; \qquad
\Delta_{ij} \ = \ d_{ij} + \omega_{ij} \quad ,
\label{delta-sum-8867}
\end{equation}
is going to be called the \emph{movement velocity}. This tensor \,$ \sfDelta $\, has no special symmetry.

%
% section
%
\section{Stress}

In this text, the symbol \,$ \sfs $\, represents the \emph{Cauchy stress tensor}, first introduced by Augustin Cauchy around 1822 (Cauchy, 1841) sometimes called the \emph{physical} stress tensor. Its components are defined via
\begin{equation}
\begin{split}
\sfs \ 
& = \ \bfS(X,t) \ = \ s^I{}_J(X,t) \ \bfe_I(X,t) \otimes \bfe^J(X,t) \\
& = \ \bfs(x,t) \ = \ s^i{}_j(x,t) \ \bfe_i(x) \otimes \bfe^j(x) \quad . \\
\end{split}
\end{equation}
In a theory like this one, where the term {\sl tensor} is used in a very restrictive sense, other matrices of numbers have no place, as, for instance, the different Piola-Kirchhoff stress ``tensors'' (Truesdell and Toupin, 1960; Eringen, 1962; Malvern, 1969; Marsden and Hughes, 1983).

In the general theory developed here is is \emph{not} assumed that the stress tensor is symmetric. So, generally, \,$ s_{IJ}\, \neq \ s_{JI} $\,, and  \,$ s_{ij}\, \neq \ s_{ji} $\,.

%
% section
%
\section{Viscosity}
\label{Viscosity}

Having introduced the stress tensor \,$ \sfs $\, and the movement velocity tensor~\,$ \sfDelta $\,, we can introduce the notion of linear viscosity by just assuming proportionality between the two:
\begin{equation}
\sfs \ = \ \sfSigma : \sfDelta \quad .
\end{equation}
Here, \,$ \sfSigma $\, is the viscosity tensor, a positive definite tensor with some symmetries\footnote{The first group of two indices and the second group of two indices can be permuted.}. I am not going to further develop here the linear theory of viscosity.

%
% section
%
\section{Elasticity}

%
% section
%
\subsection{Dealing with two temporal variables}

Some of the ``tensor functions'' to be introduced below are defined with respect to some reference time, that we will denote \,$ t_0 $\,. To denote such tensor functions, we shall use the notation \,$ {\sf s}(X,t;t_0) $\, (note the ``\,;\,''), this meaning that \,$ t_0 $\, is considered to be a fixed constant. In particular, no time derivatives can be considered with respect to \,$ t_0 $\,. As in material coordinates, the natural basis is time dependent, it will always be considered that the vector basis to be used is that at \,$ t $\,. For instance, the component of a tensor \,$ \sft $\, as the deformation or the strain tensor are \,$ t^i{}_j(x,t;t_0) $\, in the laboratory coordinates and \,$ t^I{}_J(X,t;t_0) $\, in the material coordinates, in the precise sense that one has
\begin{equation}
\begin{split}
\sft \ 
& = \ t^i{}_j(x,t;t_0) \ \bfe_i(x) \otimes \bfe^j(x) \hskip 26.8 mm x \ = \ \phi(X,t) \\
& = \ t^I{}_J(X,t;t_0) \ \bfe_I(X,t) \otimes \bfe^J(X,t) \qquad \ \qquad X \ = \ \Phi(x,t) \quad . \\
\end{split}
\end{equation}
With this convention, objects like the deformation tensor or the strain tensor (that depend on the parameter \,$ t_0 $\,) are ordinary tensors. In particular, from the relations (space variables omitted)
\begin{equation}
\bfe_i \ = \ \lambda^I{}_i(t) \, \bfe_I(t) \qquad ; \qquad
\bfe_I(t) \ = \ \lambda^i{}_I(t) \, \bfe_i \quad ,
\end{equation}
the usual rule for the change of components under a change of coordinates follows (space variables omitted):
\begin{equation}
\begin{split}
t_i{}^j(t;t_0) \ & = \ \lambda^I{}_i(t) \ t_I{}^J(t;t_0) \ \lambda^j{}_J(t) \\
t_I{}^J(t;t_0) \ & = \ \lambda^i{}_I(t) \ t_i{}^j(t;t_0) \ \lambda^J{}_j(t) \quad . \\
\end{split}
\label{tensor_rule_laboratory_material}
\end{equation}

%
% section
%
\subsection{Deformation}

I shall now introduce a tensor denoted using the symbol \,$ \sfF $\,. Its components in the laboratory and the material coordinates are defined via
\begin{equation}
\begin{split}
\sfF \ 
& = \ F_i{}^j(x,t) \ \bfe^i(x) \otimes \bfe_j(x) \hskip 26.5 mm x \ = \ \phi(X,t)\\
& = \ F_I{}^J(X,t) \ \bfe^I(X,t) \otimes \bfe_J(X,t) \qquad \ \qquad X \ = \ \Phi(x,t) \quad . \\
\end{split}
\end{equation}
There are two equivalent definitions, one using the laboratory coordinates, and one using the material coordinates (space variables omitted):
\begin{equation}
F_i{}^j(t;t_0) \ = \ \lambda^j{}_K(t) \ \lambda^K{}_i(t_0) \qquad ; \qquad
F_I{}^J(t;t_0) \ = \ \lambda^k{}_I(t) \ \lambda^J{}_k(t_0) \quad . 
\end{equation}
These two definitions are equivalent in the sense that they satisfy the tensor rule expressed by equation~\eqref{tensor_rule_laboratory_material}.

The \emph{tensor} \,$ \sfF $\, so introduced is the proper tensor replacement for the \emph{deformation gradient} ``tensor'' of the usual theory. We shall acall \,$ \sfF $\, the \emph{deformation gradient tensor} (without the quotation marks). There is no risk of confusion with the common deformation gradient, as, while that object has mixed indices, like in \,$ F_i{}^J $\,, our deformation gradient tensor always has indices \,$ F_i{}^j $\, (in laboratory coordinates) or \,$ F_I{}^J $\, (in material coordinates).

We now need to make a digression. While sometimes the symbol ``transpose'' is used by analogy with matrix theory, we need here to be precise. In particular, we need to carefully define the adjoint \,$ ^\ast \sft $\, of a tensor \,$ \sft $\,, as this is done in appendix~\ref{Transpose and adjoint}. Applying this general definition to our present problem, where we have two coordinate systems ---the laboratory one and the material one--- we find the two relations
\begin{equation}
^\ast t^i{}_j(t;t_0) \ = \ g^{ik} \ t_k{}^\ell(t;t_0) \ g_{\ell j} \quad ; \quad
^\ast t^I{}_J(t;t_0) \ = \ g^{IK}(t) \ t_K{}^L(t;t_0) \ g_{LJ}(t) \ \ . 
\end{equation}
These two definitions are equivalent in the sense that they satisfy the tensor rule expressed by equation~\eqref{tensor_rule_laboratory_material}.

We can now introduce a fundamental tensor of deformation theory, that we shall call the \emph{squared deformation tensor}:
\begin{equation}
\sfC \ = \ ^\ast \sfF \ \sfF \quad . 
\end{equation} 
Explicitly, this is
\begin{equation}
\begin{split}
C_i{}^j(t;t_0)  \ & = \ g_{ik} \ F_\ell{}^k(t;t_0) \ g^{\ell m} \ F_m{}^j(t;t_0) \\
C_I{}^J(t;t_0)  \ & = \ g_{IK}(t) \ F_L{}^K(t;t_0) \ g^{LM}(t) \ F_M{}^J(t;t_0) \\
\end{split}
\end{equation}
These two definitions are equivalent in the sense that they satisfy the tensor rule expressed by equation~\eqref{tensor_rule_laboratory_material}.
This tensor \,$ \sfC $\, satisfies the following properties:

\bigskip

{\bf Property \#1:} 
\emph{
The squared deformation tensor is \emph{self-adjoint}, i.e., one has
\begin{equation}
^\ast \sfC  \ = \ \sfC \quad .
\end{equation}
}%
(I leave to the reader to express this equation in both, laboratory and material coordinates.)

\bigskip

{\bf Property \#2:} 
\emph{
The squared deformation tensor is \emph{symmetric}, i.e., when defining
\begin{equation}
C_{ij}(t;t_0) \ = \ C_i{}^k(t;t_0) \ g_{kj} \quad ; \quad
C_{IJ}(t;t_0) \ = \ C_I{}^K(t;t_0) \ g_{KJ}(t) \ \ ,
\end{equation}
one has
\begin{equation}
C_{ij}(t;t_0) \ = \ C_{ji}(t;t_0) \quad ; \quad
C_{IJ}(t;t_0) \ = \ C_{JI}(t;t_0) \ \ .
\end{equation}
}

\medskip

{\bf Property \#3:} 
\emph{
In material coordinates, the components of the squared deformation tensor can be expressed as (making explicit the space variable~\,$ X $\,)
\begin{equation}
\boxed{ \qquad 
C_I{}^J(X,t;t_0) \ = \ g_{IK}(X,t) \, g^{KJ}(X,t_0)
\quad . \quad }
\label{gabsghat}
\end{equation}
In the laboratory coordinates, no special simplification occurs, so one just has
\begin{equation}
C_i{}^j(t;t_0) \ = \ \lambda^I{}_i(t) \ C_I{}^J(t;t_0) \ \lambda^j{}_J(t) \quad .
\end{equation}
}%

These three properties are easily demonstrated, via direct substitution.

\medskip

While the components of the squared deformation tensor \,$ \sfC $\, in the material coordinates, \,$ C_I{}^J $\,, correspond to the usual definition of the right Cau\-chy-Green deformation ``tensor'', the components in the laboratory coordinates, \,$ C_i{}^j $\,, correspond to the usual definition of the left Cauchy-Green deformation ``tensor''. While in the conventional theory two different names (right- and left- deformation tensor) are used, as well as two different symbols (usually \,$ C $\ and \,$ B $\,), we see that, in reality, there is only one tensor (with, of course, different components in different bases).
The deformation tensor originally introduced by Cauchy (in 1828) is, in fact, the inverse of our squared deformation tensor \,$ \sfC $\,.

In a work like this one, it is out of question to give different names to a unique tensor, so we have to use a single name for the tensor \,$ \sfC $\,. As in a one-dimensional elongation problem, the determinant of this tensor is
\begin{equation}
\det \sfC \ = \ ( \ell(t)/\ell(t_0) )^2 \quad , 
\end{equation}
it seems that the name here used (squared deformation tensor) is adequate.

\bigskip

{\bf Property \#4:} 
\emph{Using~\eqref{Jeroen formula 1234} and~\eqref{gabsghat}, one immediately obtains 
\begin{equation}
\dot{C}_I{}^K(X,t;t_0) \ \overline{C}_K{}^J(X,t;t_0) \ = \ 2 \, d_I{}^J(X,t) \quad ,
\label{HotHotel_001}
\end{equation}
where a ``dot'' denotes partial time derivative, and where the \,$ \overline{C}_I{}^J $\, denote the components of the tensor \,$ \sfC^{\text{-}1} $\,.}

\bigskip

The square root of the squared deformation tensor,
\begin{equation}
\boxed{ \qquad 
\sfD \ = \ \sqrt{ \, \sfC \, } \qquad} 
\end{equation}
shall naturally be named the \emph{deformation tensor}. Is is also symmetric and self-adjoint.

\bigskip

{\bf Property \#5:} 
\emph{From equation~\eqref{HotHotel_001} it follows (using the fact that \,$ \sfd $\, and \,$ \sfC $\, are symmetric)
\begin{equation}
\boxed{ \qquad 
\dot{D}_I{}^K(X,t;t_0) \ \overline{D}_K{}^J(X,t;t_0) \ = \ d_I{}^J(X,t)  \quad . \quad }
\label{HotHotel_0001}
\end{equation}
}

The expression in equation~\eqref{HotHotel_0001} mathematically corresponds to the notion of \emph{declinative} (Tarantola, 2006), that is the proper time derivative to be introduced for this kind of tensors\footnote{I am reluctantly using the name \emph{tensor} here (see footnote~\ref{test}).}: \emph{the declinative of the deformation is the deformation velocity}.

%
% section
%
%\subsection{Relation between deformation and deformation velocity}
%\label{Relation deformation tensor -- deformation velocity tensor}

{\bf Property \#6:}
\emph{From relation~\eqref{HotHotel_0001} it follows, using the matricant theory (see appendix~\ref{The matricant}), that one has (variable \,$ X $\, implicit)
\begin{equation}
D_I{}^J(t;t_0) \ = \ \delta_I{}^J + \int_{t_0}^t \!\!\! dt' \ d_I{}^J(t') + \int_{t_0}^t \!\!\! dt' \ d_I{}^K(t') \int_{t_0}^{t'} \!\!\! dt'' \ d_K{}^J(t'') + \dots \ \ .
\label{calor_utrecht}
\end{equation}
}

So, while property \#5 says that the deformation velocity is a ``properly defined'' time derivative of the deformation tensor, this property \#6 gives the inverse relation, expressing the deformation tensor as a ``properly defined'' time integral of the deformation velocity tensor. In section~\ref{Configuration space} we shall see that this is an integration of a Lie group manifold (representing the configuration space), with continuous parallel transport to the origin of the group.

Properties \#5 and \#6, taken together, suggest that our deformation tensor \,$ \sfD $\, is intimately connected to the deformation velocity tensor \,$ \sfd $\,. The deformation tensor \,$ \sfD $\, is, therefore, a fundamental tensor in the theory of continuous media.

\bigskip

{\bf Property \#7:}
\emph{One has\footnote{See appendix~\ref{The matricant}.}}
\begin{equation}
\det \bfD(X,t;t_0) \ = \ 
\exp \!\int_{t_0}^t dt' \ \text{trace} \, \bfd(X,t') \quad .
\label{ca cadra994}
\end{equation}
As \,$ \det \bfD $\, expresses the ratio between final and initial volumes, this relation relates that ratio to the time integral to the trace of the deformation velocity tensor.

%
% section
%
\subsection{Symmetric strain}
\label{Symmetric strain}

Cauchy originally defined the \emph{strain} as
\begin{equation}
\sfE \ = \ \tfrac{1}{2} ( \, \sfC  - {\sf I} \, ) \quad ,
\end{equation}
but many lines of thought suggest that this was just a guess, that, in reality, is just the first order approximation to the more proper definition
\begin{equation}
\sfE \ = \ \log\sqrt{ \, \sfC  \, } \ = \ \tfrac{1}{2} ( \, \sfC  - {\sf I} \, ) - \tfrac{1}{4} ( \, \sfC  - {\sf I} \, )^2 + \dots \quad ,
\end{equation}
i.e., in reality,
\begin{equation}
\sfE \ = \ \log\sfD \ = \  ( \, \sfD  - {\sf I} \, ) - \tfrac{1}{2} ( \, \sfD  - {\sf I} \, )^2 + \dots \quad .
\end{equation}
But this requires some care, as the logarithm of a real matrix is not always real.

\bigskip

{\bf Definition (symmetric strain):} 
\emph{Let be \,$ \sfD $\, a (symmetric) tensor\footnote{Or, if the reader pre\-fers, the \emph{matrix} representing the covariant-contravatiant components of the tensor in some basis.} that belongs to the part of the Lie group manifold \,GL$^+$(3)\, that is geodesically connected to the origin of the group. Then (Tarantola, 2006), the logarithm of \,$ \sfD $\, is a real tensor\footnote{I.e., the matrix with the components is real.}, and the \emph{(symmetric) strain} associated to \,$ \sfD $\, is defined as}
\begin{equation}
\boxed{ \qquad 
\sfE \ = \ \log \sfD  \quad . \quad }
\label{HoteHot-8833}
\end{equation}

\medskip

The reason for the strain being not defined for an arbitrary \,$ \sfD $\, is explained in section~\ref{Configuration space}. The strain defined logarithmically is often named \emph{natural strain} of \emph{Hencky strain} (e.g., Truesdell and Toupin (1960), Roug\'ee (1997)).

The components of the strain tensor are, of course, defined via
\begin{equation}
\begin{split}
\sfE \ 
& = \ E_i{}^j(x,t;t_0) \ {\bf e}^i(x) \otimes  {\bf e}_j(x) \hskip 27 mm x \ = \ \phi(X,t) \\
& = \ E_I{}^J(X,t;t_0) \ {\bf e}^I(X,t) \otimes  {\bf e}_J(X,t) \qquad \ \qquad X \ = \ \Phi(x,t) \quad . \\
\end{split}
\end{equation}
This is a bona-fide tensor. It is easy to see that this strain tensor is both, symmetric and self-adjoint.

An actual computation of the symmetric strain can be done in both, the laboratory and the material coordinates. First, one may use the property, 
\begin{equation}
\sfE \ = \ \log\sqrt{ \, \sfC \, } \ = \ \frac{1}{2} \log \sfC  \quad ,
\end{equation}
so one does not have to care about the square-root. Then, computing the logarithm of a tensor just amounts to compute the logarithm of a matrix whose entries are the mixed components (i.e., covariant-contravariant or contravariant-covariant) of the tensor in any basis. The result so obtained is intrinsic (i.e., independent from the basis being used)\footnote{This follows directly from the property that, for any invertible matrix \,$ \bfM $\,, and for any matrix \,$ \bfC $\,, one has
\,$ \bfM \ (\log \bfC) \ \bfM^{-1} \, = \, \log( \bfM \ \bfC \ \bfM^{-1} ) $\,.}.

There are different ways for computing the logarithm of a second-rank tensor given its mixed components. These range from the series expansion\footnote{
It may be that the expansion
\,$ (\log \bfC)_a{}^b \, = \, (C_a{}^b - \delta_a{}^b) - \tfrac{1}{2} \, (C_a{}^c - \delta_a{}^c) \, (C_c{}^b - \delta_c{}^b) + \tfrac{1}{3} \, (C_a{}^c - \delta_a{}^c) \, (C_c{}^d - \delta_c{}^d) \, (C_d{}^b - \delta_d{}^b) - \dots $\, 
converges. This, of course, is nothing but
\,$ \log \bfC \, = \, (\bfC - \bfI) - \tfrac{1}{2} \, (\bfC - \bfI)^2 + \tfrac{1}{3} \, (\bfC - \bfI)^3 - \dots $\,.}
to the fully analytical methods proposed in Tarantola (2006).

%
% section
%
\subsection{Asymmetric strain}
\label{Asymmetric strain}

In equation~\eqref{delta-sum-8867} I have introduced the movement velocity \,$ \sfDelta $\, as the sum of the (symmetric) deformation velocity and the (antisymmetric) rotation velocity:
\begin{equation}
\Delta_{IJ} \ = \ d_{IJ} + \omega_{IJ} \qquad ; \qquad
\Delta_{ij} \ = \ d_{ij} + \omega_{ij} \quad .
\end{equation}

To introduce the notion of an asymmetric strain, we just need to collect some of the equation above, and drop the assumption that tensors are symmetric.

The symmetric deformation tensor \,$ \sfD $\, generalizes into the asymetric deformation tensor \,$ \sfA $\, that bears with \,$ \sfDelta $\,, the same relation that \,$ \sfD $\, bears with~\,$ \sfd $\,. The equivalent of equation~\eqref{HotHotel_0001} is
\begin{equation}
\boxed{ \qquad 
\dot{A}_I{}^K(X,t;t_0) \ \overline{A}_K{}^J(X,t;t_0) \ = \ \Delta_I{}^J(X,t)  \quad , \quad }
\end{equation}
while the equivalent of the relation~\eqref{calor_utrecht} is
\begin{equation}
A_I{}^J(t;t_0) \ = \ \delta_I{}^J + \int_{t_0}^t \!\!\! dt' \ \Delta_I{}^J(t') + \int_{t_0}^t \!\!\! dt' \ \Delta_I{}^K(t') \int_{t_0}^{t'} \!\!\! dt'' \ \Delta_K{}^J(t'') + \dots \ \ .
\label{matricant series final one}
\end{equation}
The components of the tensor \,$ \sfA $\, in the laboratory coordinates are to be obtained via the usual relation implied by a change of coordinates: \,$ A_i{}^j \, = \, \lambda^I{}_i \, \lambda^j{}_J \, A_I{}^J $\,.

The equation defining the asymmetric strain is just the equivalent of equation~\eqref{HoteHot-8833}:
\begin{equation}
\boxed{ \qquad 
\sfE \ = \ \log \sfA \quad . \quad }
\end{equation}
We could use a different symbol for the asymmetric strain, but as this is just an ``obvious'' generalization, let us keep the same symbol \,$ \sfE $\,.
As above, the strain is only defined if \,$ \sfA $\, belongs to the part of the Lie group manifold \,GL$^+$(3)\, that is geodesically connected to the origin of the group, i.e., in fact, if \,$ \log\sfA $\, is real.

In the situation where there are no micro-rotations, \,$ \sfomega \, = \, 0 $\,, so \,$ \sfDelta \, = \, \sfd $\,, and \,$ \Delta $\, is symmetric. This is obviously the special case analyzed in section~\ref{Symmetric strain} (symmetric strain), and we do not need to return to it.

Let us then analyze the other extreme situation, where there are only micro-rotations. Then, the  (symmetric) deformation velocity \,$ \sfd $\, is zero, \,$ \sfDelta \, = \, \sfomega $\,, and \,$ \Delta $\, is antisymmetric. The matricant series~\eqref{matricant series final one} then gives \,$ R_I{}^J(X,t;t_0) $\, the (orthogonal) operator representing the total rotation%
\footnote{I don't know of any text there this relation between an instanteous rotation velocity \,$ \bfomega(t) $\, and the associated finite rotation \,$ \bfR(t;t_0) $\, is given, other than my own {\sl Elements for Physics} (Tarantola, 2006).}
between \,$ t_0 $\, and \,$ t $\,:
\begin{equation}
R_I{}^J(t;t_0) \ = \ \delta_I{}^J + \int_{t_0}^t \!\!\! dt' \ \omega_I{}^J(t') + \int_{t_0}^t \!\!\! dt' \ \omega_I{}^K(t') \int_{t_0}^{t'} \!\!\! dt'' \ \omega_K{}^J(t'') + \dots \ \ .
\label{second matricant 03}
\end{equation}
Note how the usual Cosserat micro-rotation enters the scene in the theory here proposed: as a (quite complex) quantity derived ---via the matricant--- from the (more elementary) micro-rotation velocity.

%
% section
%
\subsection{Hooke's law}

During the evolution of a deforming medium, different values of the time~\,$ t $\, are considered. In elasticity, one assumes that there is some reference ``configuration'' that is kept in memory by the deforming medium. Let us assume that this is the configuration at instant \,$ t_0 $\,, and let us simplify the theory by assuming that there is no ``pre-stress'', i.e., that the stress at instant \,$ t_0 $\, is zero. To remember that special condition, let us, from now on, change our notation for the stress, and denote it \,$ \sfs(t;t_0) $\, instead of just \,$ \sfs(t) $\,. The initial condition is then
\begin{equation}
\boxed{ \qquad 
s^I{}_J(t_0;t_0) \ = \ 0 
\quad . \quad }
\end{equation}

The proper formulation of linear elasticity is in material coordinates \,$ \{X,t\} $\,, because the elastic properties of a continuous medium depend on the physico-chemical properties at every material point. I define linear elasticity as the theory one obtains when assuming (this is my version of Hooke's law) that, in material coordinates,
\begin{equation}
\boxed{ \qquad 
s^I{}_J(X,t;t_0) \ = \ c^I{}_J{}^K{}_L(X,t_0) \, E_K{}^L(X,t;t_0) 
\quad , \quad }
\label{Hoooookes}
\end{equation}
where the positive definite \emph{stiffness tensor} has the symmetry\footnote{In the symmetric theory, it also has the symmetries \,$ c_{IJKL} \, = \, c_{JIKL} \, = \, c_{IJLK} $\,.}
\begin{equation}
c_{IJKL} \ = \ c_{KLIJ} \quad . 
\end{equation}
Note that, at any instant \,$ t $\,, I write Hooke's law using the components of the stiffness tensor ``frozen'' at \,$ t_0 $\,.

In the laboratory coordinates, one has
\,$ s^i{}_j = \lambda^i{}_I \lambda^J{}_j s^I{}_J = \lambda^i{}_I \lambda^J{}_j c^I{}_J{}^K{}_L E_K{}^L = \lambda^i{}_I \, \lambda^J{}_j \, c^I{}_J{}^K{}_L \, \lambda^k{}_K \, \lambda^L{}_\ell \, E_k{}^\ell $\,, 
i.e.,
\begin{equation}
s^i{}_j(x,t;t_0) \ = \ c^i{}_j{}^k{}_\ell(x,t) \,  E_k{}^\ell(x,t;t_0) \quad , \\
\end{equation}
where
\,$ c^i{}_j{}^k{}_\ell(t) \, = \, \lambda^i{}_I(t) \, \lambda^J{}_j(t) \, \lambda^k{}_K(t) \, \lambda^L{}_\ell(t) \, c^I{}_J{}^K{}_L(t_0) $\,.

Appendix~\ref{Fourth-rank isotropic (asymmetric) tensors} analyzes the example of isotropic elasticity. It is there explained the well-known fact (e.g., Nowacki, 1986) that, while in the symmetric theory, the isotropic stiffness tensor has two invariants, in the general theory it has three. 

In material coordinates, the Hooke's law implies the relation
\begin{equation} 
\dot{s}^I{}_J(X,t;t_0) \ = \ c^I{}_J{}^K{}_L(X,t_0) \, \dot{E}_K{}^L(X,t;t_0) 
\quad ,
\end{equation}
but there is no simple relation between \,$ \dot{E}_I{}^J(t;t_0) $\, and \,$ d_I{}^J(t) $\,.

Needless to say, the theory here presented is just the mathematically simplest theory. Physical reality may suggest that the ``constants'' \,$ c^I{}_J{}^K{}_L(t_0) $\, may, in fact, be functions of the temperature, the state of deformation (or the stress), etc. Also, one may need to replace the linear relation~\eqref{Hoooookes} by a more general relation. Using, for instance, some terms of a series expansion would lead to
\begin{equation}
\begin{split}
s^I{}_J(X,t;t_0) \ & = \ c^I{}_J{}^K{}_L(X,t_0) \, E_K{}^L(X,t;t_0) \\
& \hskip 4.5 mm + d^I{}_J{}^K{}_L{}^M{}_N(X,t_0) \, E_K{}^L(X,t;t_0) \, E_M{}^N(X,t;t_0) + \dots\quad , \\
\end{split}
\end{equation}
$ \sfd \, , \, \dots $\, being appropriately defined tensors.
%
% section
%
\section{Geodesic movements}
\label{Geodesic movements}

We shall say that a movement is \emph{geodesic}\footnote{We see in section~\ref{Configuration space} that such a movement actually corresponds to a geodesic path in the configuration space.} if the material components of the deformation velocity tensor are not time-dependent, \,$ d_I{}^J(X,t) \, = \, d_I{}^J(X) $\,. Then the matricant series~\eqref{matricant series final one} just becomes the exponential function, and one has
\begin{equation}
\bfA(X,t;t_0) \ = \ \exp \bfQ(X,t;t_0) \quad ,
\end{equation}
where \,$ \bfQ(X,t;t_0) $\, is the tensor whose components in the material coordinates are
\begin{equation}
Q_I{}^J(X,t;t_0) \ = \ (t-t_0) \ \, d_I{}^J(X) \quad .
\end{equation}
The strain being \,$ \sfE \, = \, \log \sfA $\,, one then has \,$ E_I{}^J(X,t;t_0) \, = \, Q_I{}^J(X,t;t_0) $\,, i.e.,
\begin{equation}
\boxed{ \qquad 
E_I{}^J(X,t;t_0) \ = \ (t-t_0) \ \, d_I{}^J(X) \quad . \quad }
\label{propstrain}
\end{equation}
So, \emph{in a geodesic movement, the components of the strain are just proportional to the components of the deformation velocity.} In particular, one has
\begin{equation}
\dot{E}_I{}^J(X,t;t_0) \ = \ d_I{}^J(X) \quad .
\end{equation}
Note that this identity between strain rate and deformation velocity is only valid for geodesic movements.

In the laboratory coordinates,
\,$ E_i{}^j \, = \, \lambda^I{}_i \, \lambda^j{}_J \, E_I{}^J \, = \, (t-t_0) \, \lambda^I{}_i \, \lambda^j{}_J \, d_I{}^J $\,, i.e.,
\begin{equation}
E_i{}^j(x,t;t_0) 
\ = \ (t-t_0) \ \, d_i{}^j(x,t) \quad .
\end{equation}
where
\,$ d_i{}^j(t) \, = \, \lambda^I{}_i(t) \, \lambda^j{}_J(t) \, d_I{}^J $\,.
Note that even if the movement is geo\-desic, the components of the deformation velocity in the laboratory coordinates are time-dependent.

%
% section
%
\section{Power and energy}

The volumetric power produced by the causes of the motion is
\begin{equation}
w(X,t) \ = \ s^I{}_J(X,t) \ d_I{}^J(X,t) \quad ,
\label{power_882245}
\end{equation}
and the volumetric energy cumulated between instant \,$ t_0 $\, and instant \,$ t $\, is
\begin{equation}
u(X,t;t_0) \ = \ \int_{t_0}^t \text{d}t' \ w(X,t') \quad .
\end{equation}

Let us first examine the case of geodesic movements (section~\ref{Geodesic movements}). Then,
using Hooke's law~\eqref{Hoooookes} and the geodesic strain relation~\eqref{propstrain}, one arrives at
\begin{equation}
w(X,t) \ = \ (t-t_0) \ \, c^I{}_J{}^K{}_L(X,t_0) \ d_K{}^L(X) \ d_I{}^J(X) \quad .
\end{equation}
Therefore,
\begin{equation}
u(t;t_0) \ = \ \frac{(t-t_0)^2}{2} \, c^I{}_J{}^K{}_L(t_0) \, d_I{}^J  \, d_K{}^L \quad ,
\end{equation}
a relation that, using again expression~\eqref{propstrain} for the geodesic strain, can be written
\begin{equation}
\boxed{ \qquad 
u(t;t_0) \ = \ \tfrac{1}{2} \, c^I{}_J{}^K{}_L(t_0) \, E_I{}^J(t;t_0) \, E_K{}^L(t;t_0) 
\quad . \quad }
\label{geodesic energy}
\end{equation}
So, \emph{for geodesic movements, the volumetric elastic energy is a quadratic function of the strain.}

The obvious question is: when the movement is not geodesic, does this elastic energy depend on the path of the movement in the configuration space? Should the answer be negative, then expression~\eqref{geodesic energy} would have general validity. This is an open question, whose answer will require to further extend the already known properties of the matricant.

%
% section
%
\section{Configuration space}
\label{Configuration space}

The configuration space of a deforming elastic medium, identified with \,GL$^+$(3)\,, was introduced with detail in Tarantola (2006), where some pictorial representations, similar to those in figures~\ref{Config_01.png}--\ref{Config_03.png} are presented.

Every (asymmetric) deformation tensor \,$ \sfA $\, (as introduced in section~\ref{Asymmetric strain}) corresponds to a point of \,GL$^+$(3)\,, so a general movement \,$ \sfA(t) $\, is an arbitrary path in  \,GL$^+$(3)\,. 

Let me now present some basic notions on the geometry of the Lie group manifold \,{\rm GL($n$)}\,. It is well-known that Lie group manifolds have a connection and a metric. The connection is not symmetric, so it is equivalent to say that Lie group manifolds have a metric and a (Cartan's) torsion. In reality, the torsion of a Lie group manifold is totally antisymmetric, this implying that geodesic lines and autoparallel lines coincide. Therefore, one can limit oneself to talk about geodesics. It is also well-known that not all the points of \,{\rm GL($n$)}\, can be reached geodesically from the origin. 

%
% figure
%
\begin{figure}[htbp]
   \centering
   \includegraphics[width=80 mm]{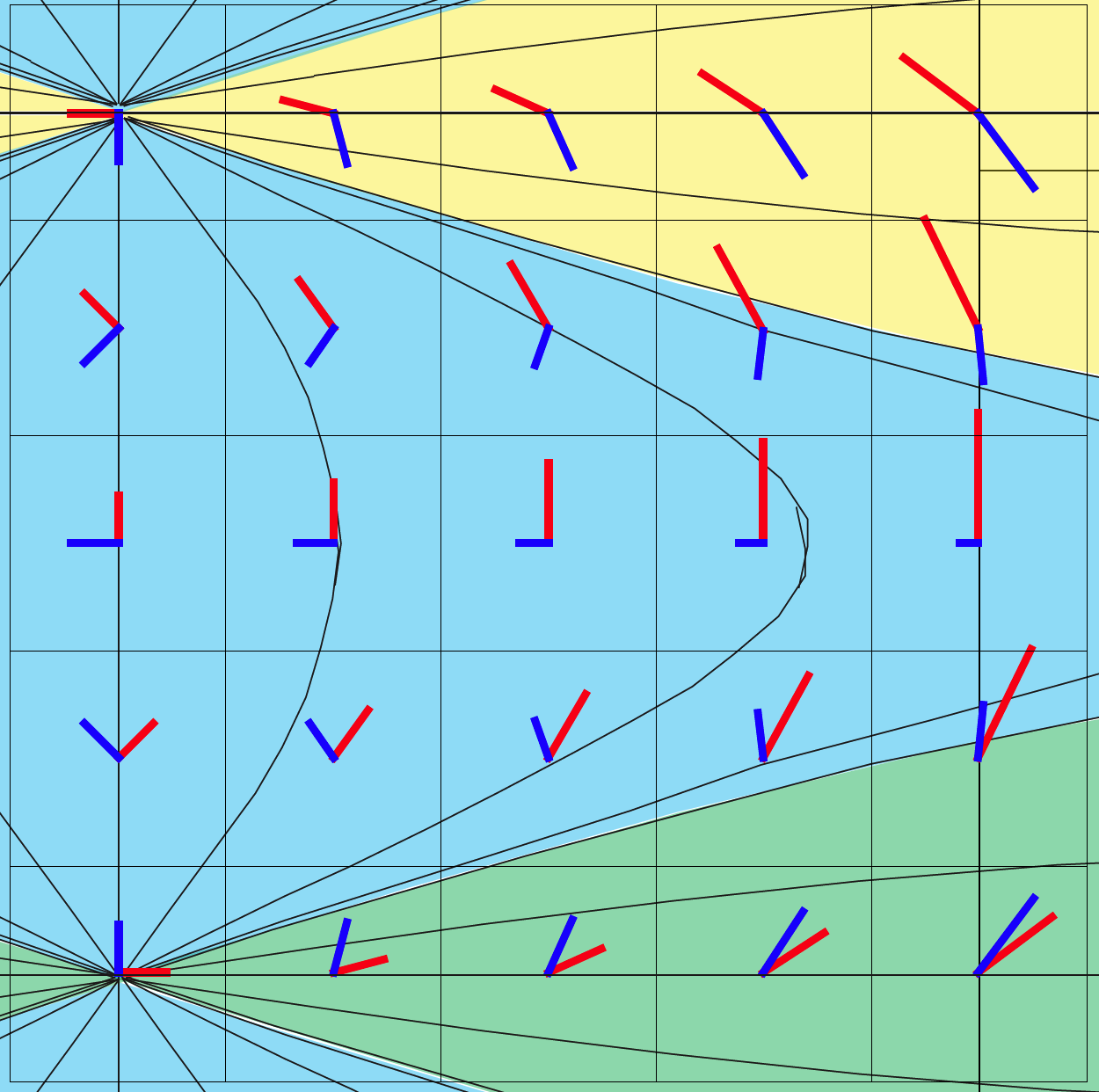} 
   \caption{In two-dimensional elasticity, the configuration space is a part of \,GL$^+$(2), and, if the deformations are volume-preserving, a part of \,SL(2)\,. A~(partial) section of \,SL(2)\, is represented here, together with some its geodesics (thin black lines). See Tarantola (2006) for details. Each point of \,SL(2)\, can be seen as a transformation: that transforming one vector basis into another. Here, the basis at the bottom-left (origin of the group) is transformed into the other bases represented. The points in the yellow area can no be reached geodesically from the origin. The logarithm of the matrices in the yellow area are not real (the components of the logarithm are complex numbers).}
   \label{Config_01.png}
\end{figure}

%
% figure
%
\begin{figure}[htbp]
   \centering
   \includegraphics[width=120 mm]{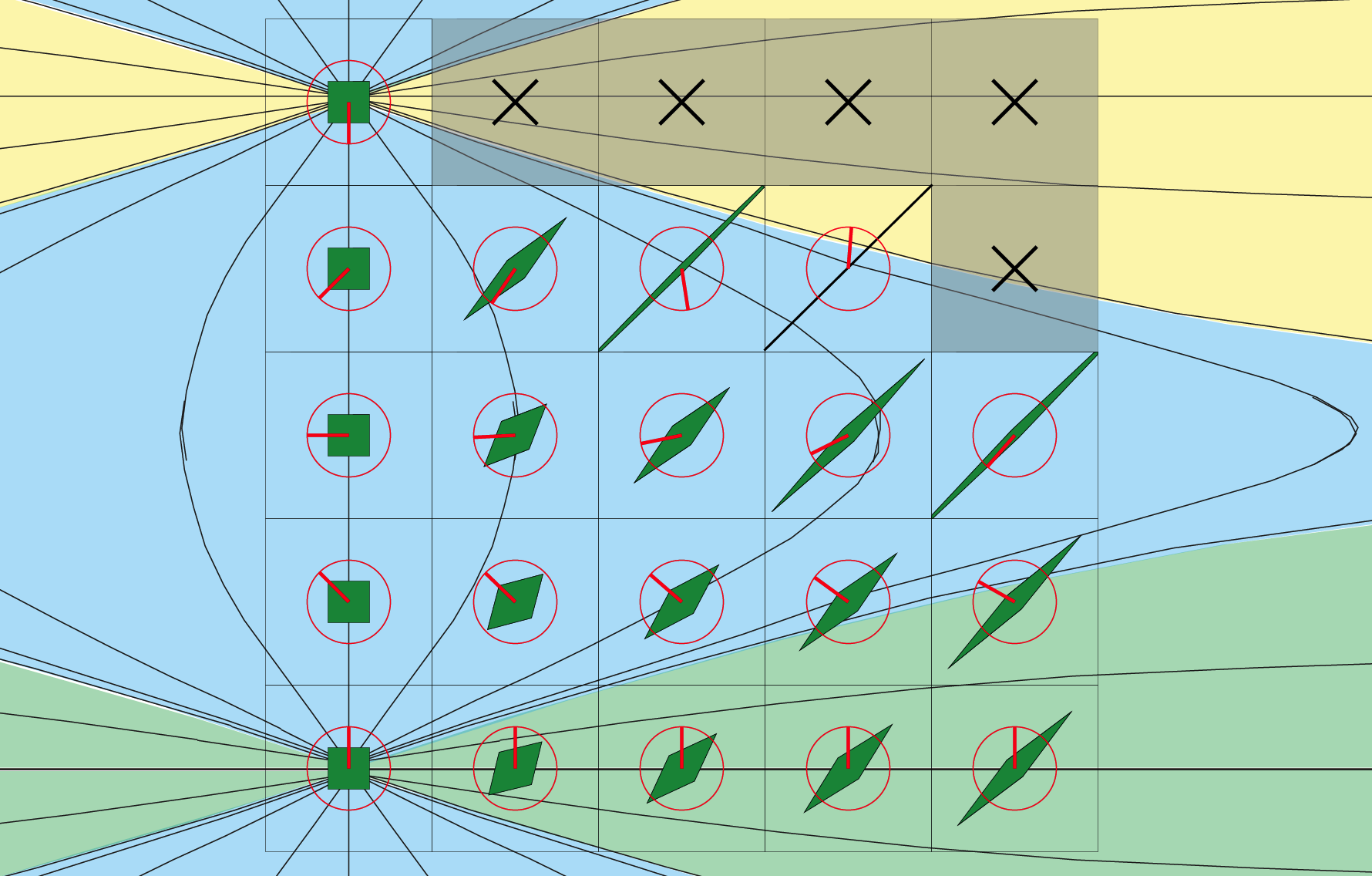} 
   \caption{In two-dimensional elasticity, the part of \,SL(2)\, that is geodesically connected to the origin is made of the matrices whose logarithm is real. Let \,$ \bfA $\, be such a matrix. It can be identified to the deformation tensor of the text. Then, the strain \,$ \bfE \, = \, \log \bfA $\, is defined (it is real). The symmetric part of the strain represents a deformation, and the antisymmetric part, a (Cosserat) rotation. Both, the deformation and the rotation are represented in this figure. In the grey area, points that belong to \,SL(2)\, but not to the configuration space. Of course the configuration space contains all possible deformations and rotations.}
   \label{Config_02.png}
\end{figure}

%
% figure
%
\begin{figure}[htbp]
   \centering
   \includegraphics[width=120 mm]{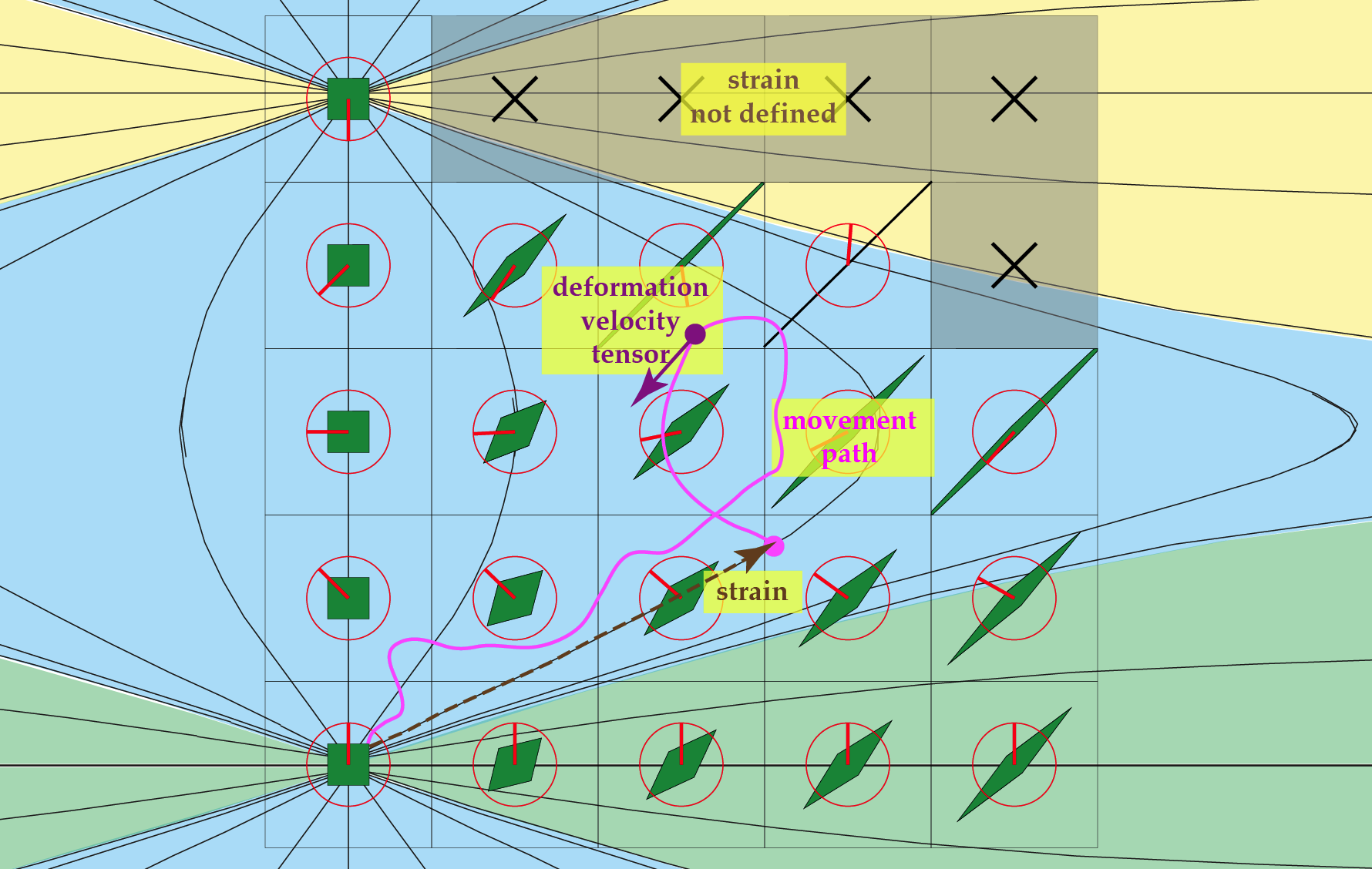} 
   \caption{Continued from figure~\ref{Config_02.png}. The curved line represents an arbitrary movement. The movement velocity \,$ \Delta $\, (denoted ``deformation velocity tensor'' in the figure), is the tangent to any point of this line (in reality, after its \emph{transport to the origin of the group}). In the terminology of Tarantola (2006), this is the declinative of the path. The final configuration of the movement is the big pink dot. The strain associated to this final configuration is the dashed geodesic segment (this is the logarithm of the \,$\bfA $\, representing the point). For the points of the configuration space not geodesically connected to the origin the strain is not defined (and the logarithm would not be real): physically, these configurations can not be accessed from the reference configuration by a stress of the form \,$ \lambda \, \sfs $\,, where \,$ \sfs $\, is a fixed (``nominal stress'') and \,$ \lambda $\, some parameter (for instance, the time \,$ t $\,) varying from zero to infinity.}
   \label{Config_03.png}
\end{figure}

A matrix \,$ \bfM $\, of \,{\rm GL($n$)}\,, has ``entries'', say \,$ M_a{}^b $\,. When choosing these \,$ n^2 $\, entries as a coordinate system on  \,{\rm GL($n$)}\,, there is the associated natural basis at every point. Tarantola(2006) demonstrates the following property: \emph{if \,$ \bfM $\, belongs to the part of  \,{\rm GL($n$)}\, that is geodesically connected to the origin, then \,$ \bfm \, = \, \log \bfM $\, is real, and the entries of \,$ \bfm $\, are the components on the natural basis at the origin \,$ \bfI $\, of the group of the vector of the tangent space (``algebra'') that is tangent to the geodesic connecting \,$ \bfI $\, to \,$ \bfM $\, (on the group) and whose norm is equal to the length of the geodesic.}
From where the logarithmic definition of the strain. 

Now, consider a point \,$ \sfA $\, of \,GL$^+$(3), that belongs to the configuration space\footnote{I.e., that belongs to the part of  \,GL$^+$(3)\, that is geodesically connected to the origin of the group.}. Then, \,$ \sfE \, = \, \log\sfA $\, is real, and can be interpreted as the (dashed) geodesic segment in figure~\ref{Config_03.png}. The decomposition of the strain into its symmetric and its antisymmetric part,
\begin{equation}
\sfE \ = \ \widetilde{\sfE} + \widehat{\sfE} \quad ,
\end{equation}
allows to consider the (symmetric) deformation tensor
\begin{equation}
\sfD \ = \ \exp \widetilde{\sfE} 
\end{equation}
---that is a ``standard'' deformation--- and the (orthogonal) rotation tensor
\begin{equation}
\sfR \ = \ \exp \widehat{\sfE} 
\end{equation}
that corresponds to the Cosserat micro-rotations. This decomposition of \,$ \sfA $\, into a (symmetric) deformation \,$ \sfD $\, and an (orthogonal) rotation \,$ \sfR $\, is at the basis of the representation of the configuration space in figures~\ref{Config_02.png} and~\ref{Config_03.png}.

%
% section
%
\section{Bibliography}

\def\bibref{\par\noindent\hangindent=20pt}

\bibref
Cauchy, A.-L., 1841,
M\'emoire sur les dilatations, les condensations et les rotations produites
par un changement de forme dans un syst\`eme de points mat\'eriels,
Oeuvres compl\`etes d'Augustin Cauchy, II--XII, pp.\ 343--377, Gauthier-Villars,
Paris.

\bibref
Coll, B., 2002, 
A principal positioning system for the Earth, 
JSR 2002, eds. N. Capitaine and M. Stavinschi,
Pub.\ Observatoire de Paris, pp.\ 34Ð-38.

\bibref
Cosserat, E.\ and F.~Cosserat, 1909,
Th\'eorie des corps d\'eformables,
A. Hermann, Paris.

\bibref
Eringen, A.C., 1962,
Nonlinear theory of continuous media,
McGraw-Hill, New York.

\bibref
Gantmacher, F.R., 1967,
Teorija matrits, Nauka, Moscow.
English translation, The theory of matrices, Chelsea Pub Co., 2000. 

\bibref
Halwachs, F., 1960, Th\'eorie relativiste des fluides \`a spin, Gauthier Villars.

\bibref
Kennett, B., 1983, Seismic wave propagation is stratified media, Cambridge University Press. Now freely available at the Australian National University electronic press ({\sc anu e press}).

\bibref
Malvern, L.E., 1969,
Introduction to the mechanics of a continuous medium,
Prentice-Hall.

\bibref
Marsden J.E., and Hughes, T.J.R., 1983,
Mathematical foundations of elasticity, 
Dover.

\bibref
Nowacki, W., 1986,
Theory of asymmetric elasticity, 
Pergamon Press.

\bibref
Peano, G., 1888, Int\'egration par s\'eries des \'equations diff\'erientielles lin\'eaires, Math.\ Ann.\ Vol.\ 32, pp.\ 450--456.

\bibref
Roug\'ee, P., 1997,
M\'ecanique des grandes transformations,
Springer.

\bibref
Sedov, L., 1973,
Mechanics of continuous media, 
Nauka, Moscow.
French translation: M\'ecanique des milieux continus, Mir, Moscou, 1975.

\bibref
Tarantola, A., 2006, Elements for Physics, Springer.

\bibref
Tarantola, A., L.\ Klime\v{s}, J.M.\ Pozo, and B.\ Coll, 2009, 
Gravimetry, relativity, and the global navigation satellite systems, 
arXiv: 0905:3798.

\bibref
Truesdell C., and Toupin, R., 1960,
The classical field theories,
{\em in:\/} Encyclopedia of physics, 
edited by S.~Fl\"ugge, 
Vol.\ III/1, 
Principles of classical mechanics and field theory,
Springer-Verlag, Berlin.

%\newpage

%
% section
%
\section{Acknowledgements}

This work started during my recent stay at Princeton University, where I~had extremely inspiring discussions wit Jeroen Tromp, Marcelo Epstein, Michel Slawinski, and Andrew Norris. The remarks of Marcelo on the properties of hypo elasticity were determinant. Jeroen and I have been intermittently working in this topic for some years now. We were not happy with present theories of finite deformation and of finite elasticity. The background of the theory here presented (a proper definition of motion and of deformation velocity) was elaborated in March 2009, while I~was at Princeton with Jeroen and Michel. A formula that, in retrospect, has proven to be fecund is the relation~\eqref{Jeroen formula 1234} expressing, in the material coordinates, the components of the deformation velocity tensor as the partial time derivative of the components of the metric tensor. It was obtained by Jeroen. When I went back to Paris, Jeroen and I tried to continue developing the theory, but a rift soon appeared: while Jeroen was insisting in a definition of the strain as a simple time integral of the deformation velocity tensor, I~was insisting on a logarithmic definition. I had a constraint that Jeroen did not accept: that the results are consistent with the geometrical developments in my book {\sl Elements for Physics}. While, at present, Jeroen considers that a quadratic dependence of the elastic energy on the strain has to be taken as an axiom, I have elaborated all my the theory around the notion of matricant (it is already present in page 227 of my {\sl Elements}, and this should have prompted me to make the link between deformation and deformation velocity there; in retrospect, I~don't understand why I didn't). The rift separating our points of view has now grown so wide, that Jeroen prefers to follow his own path, independent of mine. I regret.

%
% section
%
\section{Appendices}

%
% section
%
\subsection{The matricant}
\label{The matricant}

What follows is an exposition of the notion of \emph{matricant}, as exposed by Gantmacher (2000). I~limit myself to small adaptations of notations and of language.

Letting \,$ X(t) $\, and \,$ P(t) $\, be two time-dependent matrices, Gantmacher considers the differential matrix equation
\begin{equation}
\dot{X}(t) \, X(t)^{-1} \ = \ P(t) \quad .
\label{Gantmacher_01}
\end{equation}
Here, \,$ P(t) $\, is assumed to be ``a continuous matrix function of the argument \,$ t $\, in some interval \,$ (a,b) $\,''.
A solution to the system~\eqref{Gantmacher_01} is sought such that for some \,$ t_0 $\, in the interval \,$ (a,b) $\,, the solution satisfies \,$ X(t_0) \, = \, I $\,. Such a solution is determined by ``the method of successive approximations''. The successive approximations \,$ X_k(t) \ \ (k=0,1,2,\dots) $\, are found from the recurrence relations
\begin{equation}
\dot{X}_k(t) X_{k-1}(t)^{-1} \ = \ P(t) \qquad (k=1,2\dots) \quad ,
\label{Gantmacher_02}
\end{equation}
where \,$ X_0(t) $\, is taken equal to the identity matrix \,$ I $\,.
Setting \,$ K_k(t_0) \, = \, I \ \ (k=0,1,2,\dots) $\,, one may represent \,$ X_k(t) $\, in the form
\begin{equation}
X_k(t) \ = \ I + \int_{t_0}^t dt' \ P(t') \ X_{k-1}(t') \quad ,
\label{Gantmacher_03}
\end{equation}
thus,
\,$ X_0(t) \, = \, I $\,,
\,$ X_1(t) \, = \, I + \int_{t_0}^t dt' \ P(t') $\,,
\,$ X_2(t) \, = \, I + \int_{t_0}^t dt' \ P(t') \big( I + \int_{t_0}^{t'} dt'' \ P(t'') \big)
\, = \, I + \int_{t_0}^t dt' \ P(t') + \int_{t_0}^t dt' \ P(t') \int_{t_0}^{t'} dt'' \ P(t'') $\,,
etc.
Then, Gantmacher proves that this series,
\begin{equation}
\Omega(t;t_0) \ = \ I +  \int_{t_0}^t dt' \ P(t') 
+ \int_{t_0}^t dt' \ P(t') \int_{t_0}^{t'} dt'' \ P(t'') + \dots
\quad , 
\label{Gantmacher_04}
\end{equation}
is absolutely and uniformly convergent in every closed subinterval of the interval \,$ (a,b) $\,, and, therefore, constitutes the solution of~\eqref{Gantmacher_01}.

That the sum~\eqref{Gantmacher_04} is the solution of~\eqref{Gantmacher_01} is verified by a term-by-term differentiation. ``This term-by-term differentiation is permissible, because the series obtained after differentiation differs from~\eqref{Gantmacher_04} by the factor \,$ P(t) $\, and, therefore, like~\eqref{Gantmacher_04}, is uniformly convergent in every closed interval contained in \,$ (a,b) $\,''. As already anticipated in expression~\eqref{Gantmacher_04}, this ``normal'' solution (often called the \emph{matricant}) is denoted \,$ \Omega(t;t_0) $\,. Gantmacher explains that every other solution is of the form
\begin{equation}
X(t) \ = \ \Omega(t;t_0) \, C \quad , 
\label{Gantmacher_05}
\end{equation}
where \,$ C $\, is an arbitrary constant matrix. Gantmacher says that it follows from this formula that every solution, in particular, the normalized one, is uniquely determined by its value for \,$ t \, = \, t_0 $\,.

The representation of the matricant in the form of such a series was first obtained by Giuseppe Peano (Peano, 1888). The matricant theory is used in seismology, typically for the propagation of wave fields in depth: I first learned about the matricant when reading Brian Kennett's book (Kennett, 1983).

\medskip

Property \#1: One has
\,$ \Omega(t;t_0) \, = \, \Omega(t;t_1) \ \Omega(t_1;t_0) $\,.

\medskip

Property \#2: One has
\,$
\Omega(t;t_0)(P+Q) \, = \, \Omega(t;t_0)(P) \ \Omega(t;t_0)(S) $\,,
with
\,$
S \, = \, \Omega(t_0;t)(P) \ Q \ \Omega(t;t_0)(P) $\,.

\medskip

Property \#3: One has
\,$
\det \Omega(t;t_0) \, = \, \exp \big( \int_{t_0}^t dt' \ \text{tr} \, P(t')  \big) $\,.

\medskip

Property \#4:
If \,$ P $\, is constant,
\,$
\Omega(t;t_0) \, = \, \exp \big( (t-t_0) \, P  \big) $\,.

%
% section
%
\subsection{Transpose and adjoint}
\label{Transpose and adjoint}

Let \,$ \SpaceV $\, be a linear space, \,$ ^\ast\SpaceV $\, its dual. The mathematical definition of the dual of a linear space is abstract\footnote{It is the linear space containing all the linear forms over \,$ \SpaceV $\,.}, but we only need here the most basic of its properties: if \,$ \SpaceV $\, is a space of vectors with components \,$ v^i $\, (in some vector basis), then \,$ \SpaceX \, = \, ^\ast\SpaceV $\, is a linear space of objects (forms) with components \,$ x_i $\, (in some form basis), so that the expression 
\begin{equation}
\langle \ \bfx \, , \, \bfv \ \rangle \ \equiv \ x_i \, v^i \ \equiv \ \sum_i  x_i \, v^i 
\end{equation}
makes sense. This is called the \emph{duality product}.

A relation like
\begin{equation}
\bfw \ = \ \bfS \, \bfv \qquad ; \qquad w^\alpha\ = \ S^\alpha{}_i \, v^i
\end{equation}
defines a linear mapping from a linear space \,$ \SpaceV $\, into a linear space \,$ \SpaceW $\,. The \emph{transpose} of the mapping, denoted \,$ ^t\bfS $\,, is, by definition the linear mapping from \,$ \SpaceY \ = \ ^\ast \SpaceW $\,, the dual of \,$ \SpaceW $\,, into \,$ \SpaceX \ = \ ^\ast \SpaceV $\,, the dual of \,$ \SpaceV $\,, such that the relation
\begin{equation}
\langle \ ^t\bfS \ \bfy \, , \, \bfv \ \rangle \ = \ \langle \ \bfy \, , \, \bfS \ \bfv \ \rangle 
\end{equation}
holds in general. Explicitly, this is
\begin{equation}
(^t\bfS \, \bfy)_i \ v^i\ = \ y_\alpha \, (\bfS \, \bfv)^\alpha \quad .
\label{transpose_01}
\end{equation}
While the components of \,$ \bfS $\, were denoted using the indices \,$ S^\alpha{}_i $\, it is convenient to denote \,$ ^t\bfS $\, using the same symbol \,$ S $\,, and just changing the positions of the indices: \,$ S_i{}^\alpha $\,. The condition in equation~\eqref{transpose_01} then becomes
\begin{equation}
(S_i{}^\alpha \ y_\alpha) \ v^i\ = \ y_\alpha\, (S^\alpha{}_i \ v^i) \quad .
\label{faim-4493}
\end{equation}
It is clear that for this relation have gereral validity, one must have
\begin{equation}
S_i{}^\alpha \ = \ S^\alpha{}_i \quad ,
\end{equation}
that is the relation holding between the components of a linear mapping and its transpose. Practically, \emph{excepted for a ``replacement of the indices'' there is no difference between a mapping and its transpose.}

In the same context, assume now that the two spaces \,$ \SpaceV $\, and \,$ \SpaceW $\, are, in fact, scalar product vector spaces, i.e., assume that that there exist two metric tensors \,$ g_{ij} $\, and \,$\gamma_{\alpha\beta} $\, defining the two scalar products
\begin{equation}
( \, \bfv_1\, , \, \bfv_2 \, ) \ = \ g_{ij} \, v_1^i \, v_2^j \qquad ; \qquad
( \, \bfw_1\, , \, \bfw_2 \, ) \ = \ \gamma_{\alpha\beta} \, w_1^\alpha \, w_2^\beta \quad .
\end{equation}
Then, in addition to the transpose, one can introduce the \emph{adjoint}, denoted \,$ ^\ast\bfS $\,, that is, by definition the linear mapping from \,$ \SpaceW $\,, into \,$ \SpaceV $\,, such that the relation
\begin{equation}
( \ ^\ast\bfS \ \bfw \, , \, \bfv \ ) \ = \ ( \ \bfw \, , \, \bfS \ \bfv \ ) 
\end{equation}
holds in general. Explicitly, this is
\begin{equation}
g_{ij} \, (^\ast\bfS \, \bfw)^i \ v^j 
\ = \ \gamma_{\alpha\beta} \, w^\alpha \, (\bfS \, \bfv)^\beta
\ = \ \gamma_{\alpha\beta} \, w^\alpha \, (S^\beta{}_i \, v^i)
\quad ,
\end{equation}
i.e., using the relation~\eqref{faim-4493} involving the transpose,
\begin{equation}
g_{ij} \, (^\ast\bfS \, \bfw)^i \ v^j 
\ = \ (S_i{}^\beta \ \gamma_{\alpha\beta} \, w^\alpha) \ v^i
\quad .
\end{equation}
In order for this to hold unconditionally, one must have
\begin{equation}
(^\ast\bfS \, \bfw)^i 
\ = \ ^\ast S_\alpha{}^i \ w^\alpha
\quad ,
\end{equation}
with
\begin{equation}
^\ast S_\alpha{}^i \ = \ \gamma_{\alpha\beta} \,  S_j{}^\beta g^{ji} \quad .
\end{equation}

The situation found in the text is a special case of this, where the mapping \,$ \bfS $\, is an endomorphism (mapping a linear space into itself), so there is only one metric.

%
% section
%
\subsection{Fourth-rank isotropic (asymmetric) tensors}
\label{Fourth-rank isotropic (asymmetric) tensors}

Here, the notion of fourth-rank isotropic tensor is discussed, without any particular reference to elasticity or viscosity.

The viscuous of elastic invariants (eigenvalues of the fourth-rank isotro\-pic tensor) will typically be a function of the material point. If using the material coordinates, we shall then face the functions
\begin{equation}
\lambda_\kappa \ = \ \Gamma_\kappa(X) 
\qquad ; \qquad
\lambda_\mu \ = \ \Gamma_\mu(X) 
\qquad ; \qquad
\lambda_\omega \ = \ \Gamma_\omega(X) 
\quad ,
\end{equation}
while, if using the laboratory coordinates, we shall face the functions
\begin{equation}
\lambda_\kappa \ = \ \gamma_\kappa(x,t) 
\qquad ; \qquad
\lambda_\mu \ = \ \gamma_\mu(x,t) 
\qquad ; \qquad
\lambda_\omega \ = \ \gamma_\omega(x,t) 
\quad ,
\end{equation}
related to the previous ones via
\begin{equation}
\begin{split}
\lambda_\kappa \ & = \ \gamma_\kappa(x,t) \ = \ \Gamma_\kappa( \, \Phi(x,t) \, ) \\
\lambda_\mu \ & = \ \gamma_\mu(x,t) \ = \ \Gamma_\mu( \, \Phi(x,t) \, ) \\
\lambda_\omega \ & = \ \gamma_ \omega(x,t) \ = \ \Gamma_ \omega( \, \Phi(x,t) \, ) \quad . \\
\end{split}
\end{equation}
Note the time-dependency of the functions in the laboratory coordinates.

In material coordinates, the components of a fourth rank isotropic tensor are
\begin{equation}
\begin{split}
c^I{}_J{}^K{}_L(X,t) \ 
= \ & \Gamma_\kappa(X) \, k^I{}_J{}^K{}_L \\
  + & \Gamma_\mu(X) \, m^I{}_J{}^K{}_L(X,t) \\
  + & \Gamma_\omega(X) \, a^I{}_J{}^K{}_L(X,t) \quad ,
\end{split}
\end{equation}
with the three orthogonal projectors
\begin{equation}
\begin{split}
k^I{}_J{}^K{}_L \ & = \ \tfrac{1}{3} \, \delta^I{}_J \ \delta^K{}_L \\
m^I{}_J{}^K{}_L(X,t) \ & = \ \tfrac{1}{2} \, ( \, g^{IK}(X,t) \ g_{JL}(X,t) + \delta^I{}_L \ \delta_J{}^K \, ) - \tfrac{1}{3} \, \delta^I{}_J \ \delta^K{}_L \\
a^I{}_J{}^K{}_L(X,t) \ & = \ \tfrac{1}{2} \, ( \, g^{IK}(X,t) \ g_{JL}(X,t) - \delta^I{}_L \ \delta_J{}^K \, ) \quad . \\
\end{split}
\end{equation}

In laboratory coordinates, the components of a fourth rank isotropic tensor are
\begin{equation}
\begin{split}
c^i{}_j{}^k{}_\ell (x,t) \ 
= \ & \gamma_\kappa(x,t) \, k^i{}_j{}^k{}_\ell  \\
  + & \gamma_\mu(x,t) \, m^i{}_j{}^k{}_\ell (x) \\
  + & \gamma_\omega(x,t) \, a^i{}_j{}^k{}_\ell (x) \quad ,
\end{split}
\end{equation}
with the three orthogonal projectors
\begin{equation}
\begin{split}
k^i{}_j{}^k{}_\ell  \ & = \ \tfrac{1}{3} \, \delta^i{}_j \ \delta^k{}_\ell  \\
m^i{}_j{}^k{}_\ell (x) \ & = \ \tfrac{1}{2} \, ( \, g^{ik}(x) \ g_{j\ell}(x) + \delta^i{}_\ell  \ \delta_j{}^k \, ) - \tfrac{1}{3} \, \delta^i{}_j \ \delta^k{}_\ell  \\
a^i{}_j{}^k{}_\ell (x) \ & = \ \tfrac{1}{2} \, ( \, g^{ik}(x) \ g_{j\ell}(x) - \delta^i{}_\ell  \ \delta_j{}^k \, ) \quad . \\
\end{split}
\end{equation}

Note that while in the material coordinates, the time-dependencies are in the components of the metric, \,$ g_{IJ}(X,t) $\, and \,$ g^{IJ}(X,t) $\,, in the material coordinates they are in the functions \,$ \gamma_\kappa(x,t)$\,,
\,$ \gamma_\mu(x,t) $\,, and \,$ \gamma_\omega(x,t) $\,.

These components are related via
\begin{equation}
c^I{}_J{}^K{}_L \ = \ \lambda^I{}_i \ \lambda^j{}_J \ \lambda^K{}_k \ \lambda^\ell{}_L \ c^i{}_j{}^k{}_\ell 
\quad ; \quad
c^i{}_j{}^k{}_\ell \ = \ \lambda^i{}_I \ \lambda^J{}_j \ \lambda^k{}_K \ \lambda^L{}_\ell \ c^I{}_J{}^K{}_L 
\ \ ,
\end{equation}
as it should.

\vskip 25 mm

\centerline{---}
\centerline{\sc See the next two pages for the Tables of Formulas.}
\centerline{---}

\newpage

%
% section
%
\subsection{Tables of formulas}

\vskip 12 mm

\begin{figure}[htbp] %  figure placement: here, top, bottom, or page
   \centering
   \includegraphics[width=96 mm]{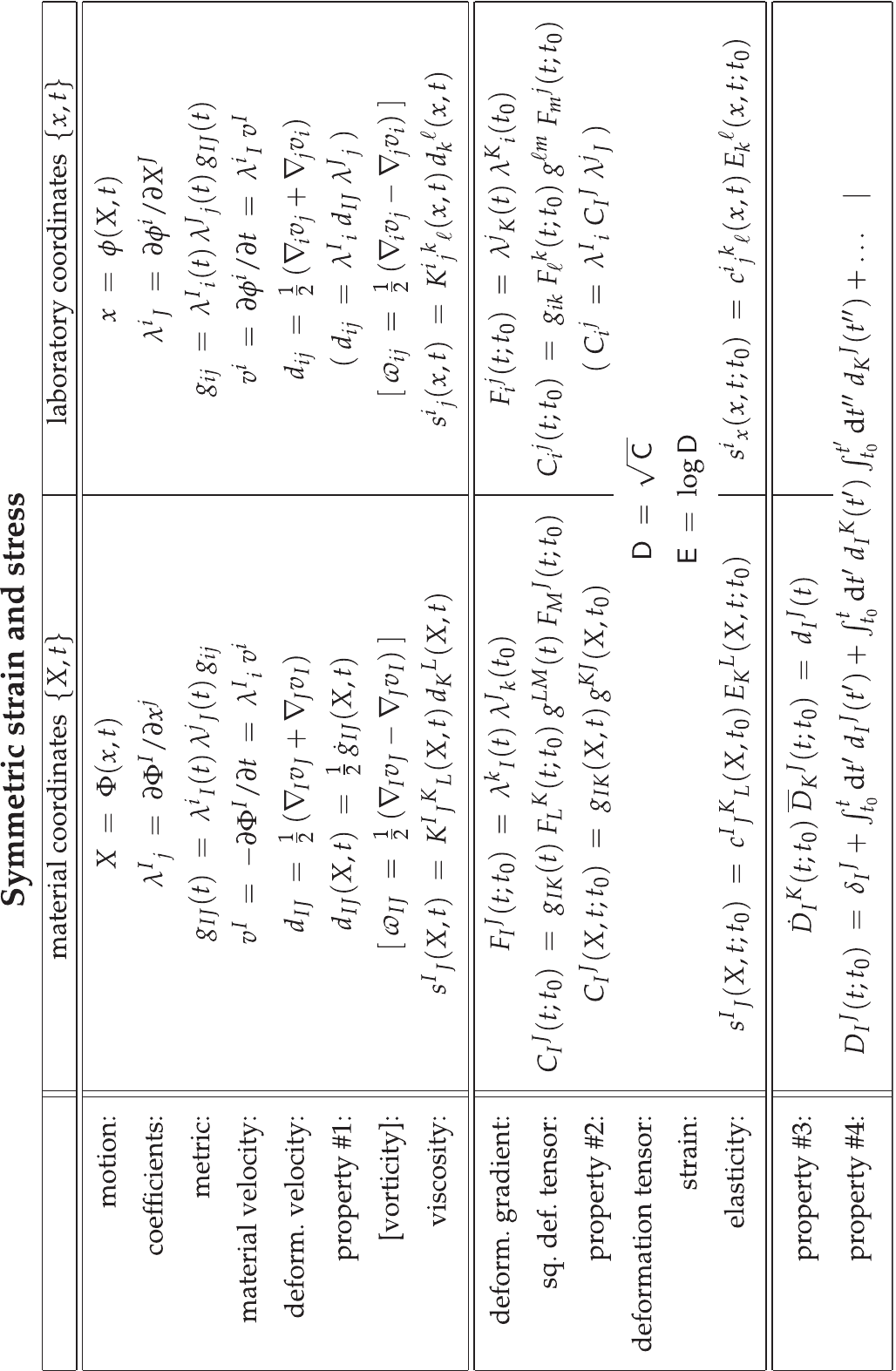} 
%   \caption{example caption}
\end{figure}

\newpage

\quad 
\vskip 10 mm

\begin{figure}[htbp] %  figure placement: here, top, bottom, or page
   \centering
   \includegraphics[width=92mm]{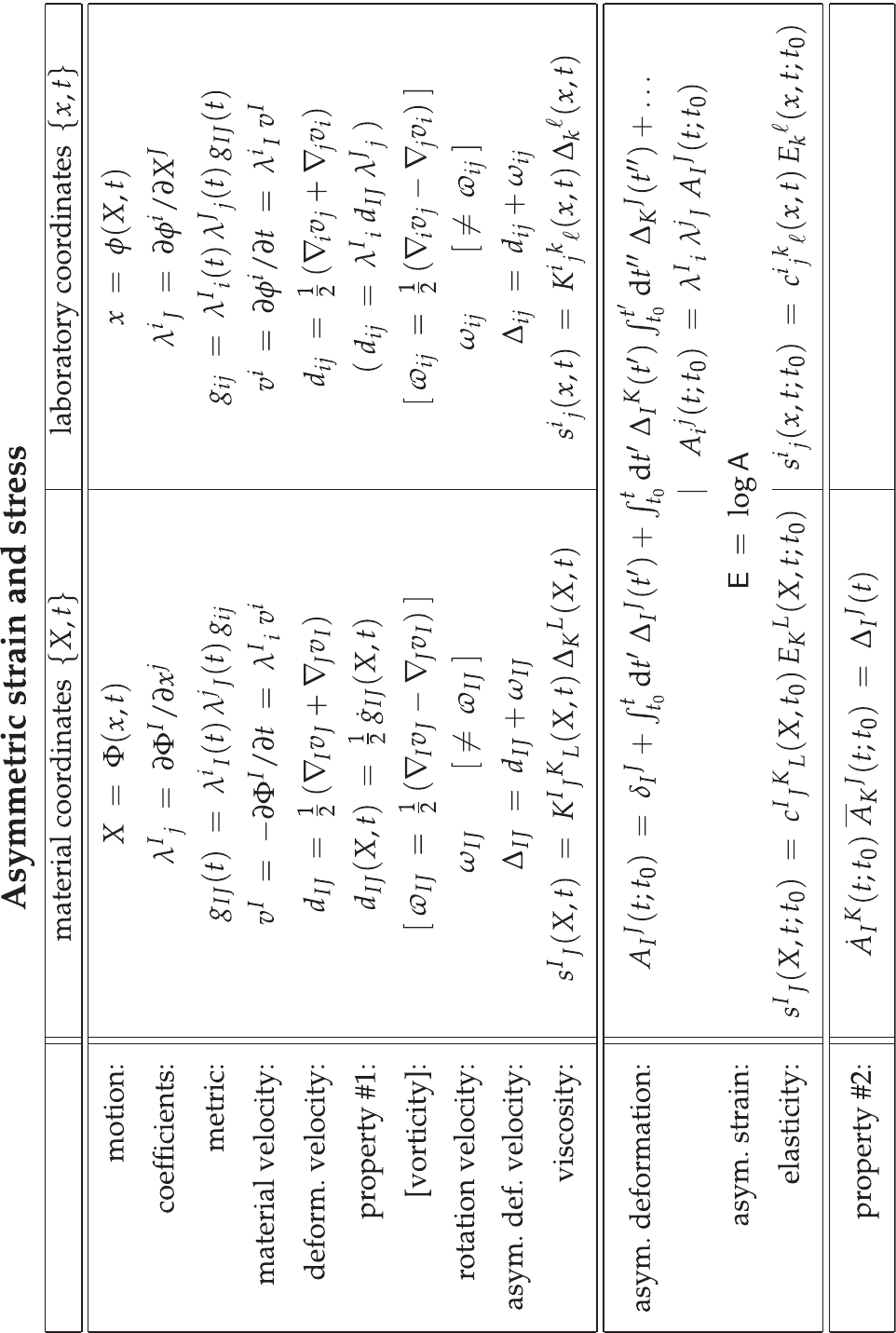} 
%   \caption{example caption}
\end{figure}

 \end{document}